\documentclass[a4paper,nobibnotes,nofootinbib]{revtex4}

\usepackage{amsmath,amssymb}
\usepackage{epsfig}

\newcommand{\g}{\gamma}
\newcommand{\f}{\frac}
\newcommand{\bea}{\begin{eqnarray}}

\newcommand{\D}{\displaystyle}

\newcommand{\intc}[1]{{\int\frac{d#1}{2i\pi}}}

\begin{document}
\title{Saturation at Hadron Colliders}

\author{C. Marquet}\email{marquet@spht.saclay.cea.fr} 
\author{R. Peschanski}
\email{pesch@spht.saclay.cea.fr}
\affiliation{Service de physique th{\'e}orique, CEA/Saclay,
  91191 Gif-sur-Yvette cedex, France\footnote{%
URA 2306, unit{\'e} de recherche associ{\'e}e au CNRS.}}

\begin{abstract}
We extend  the saturation models {\it \`{a} la} Golec-Biernat and W\"usthoff 
to cross-sections of hard  processes
initiated by virtual-gluon probes separated by large rapidity intervals at 
hadron colliders. We 
derive their analytic expressions and apply them to physical examples, such as 
saturation effects 
for  Mueller-Navelet forward jets. By comparison to $\gamma^*\!-\!\gamma^*$  
cross-sections we find a more abrupt transition to saturation. We propose to 
study observables with a potentially 
clear saturation signal and to use heavy vector and flavored mesons as
alternative hard probes to forward jets.
\end{abstract}

\maketitle

\section{Introduction}
\label{1}

The saturation regime  describes the high-density phase 
of 
partons in perturbative QCD. It may occur for instance when the 
Balitsky-Fadin-Kuraev-Lipatov (BFKL) QCD evolution equation \cite{bfkl} goes 
beyond some energy 
limit \cite{GLR,qiu,venugopalan,Balitsky:1995ub,levin,Mueller:2002zm}. On a 
phenomenological ground, a well-known saturation model \cite{golec} by 
Golec-Biernat and W\"usthoff (GBW) gives a parametrisation  of  
the proton structure functions
already in the HERA energy 
range. It   provides  a simple and elegant formulation of the transition to 
saturation. 
However, there 
does 
not yet exist a clear  confirmation of  saturation since  the same data can 
well be explained within 
the conventional  perturbative QCD framework \cite{dglap}. 

An interesting question is whether the experiments at high-energy hadron 
colliders, 
such as  the Tevatron or LHC, can test
saturation while for the moment this search is mainly considered for 
heavy-ion collisions.
In the present paper, our aim is to look for saturation effects in the 
context of Mueller-Navelet \cite{navelet} forward-jet production in 
hadron-induced hard 
collisions.  The key difference with 
electron-induced 
reactions is that the hard probe is no more a virtual photon $\g^*$ but a 
virtual gluon $g^*,$ see 
Fig.1a.

A 
basic ingredient of the GBW saturation models is the QCD dipole 
formalism \cite{niko,mueller} in which the hard  
cross-sections read
\begin{equation}
\sigma=\int d^2r_1\ d^2r_2\ \phi^{(1)}(r_1,Q^2_1) \ 
\phi^{(2)}(r_2,Q^2_2)\ \sigma_{dd}(\Delta \eta,r_1,r_2)\ ,
\label{sigma}
\end{equation}
where $r_{i=1,2}$ are the transverse sizes of the dipoles and  
$\Delta \eta$ is the pseudo-rapidity range of the dipole-dipole 
cross-section
$\sigma_{dd}(\Delta \eta,r_1,r_2).$
In our notations, $\phi^{(i)}(r_i,Q^2_i)$ are the  dipole distributions 
in 
the 
target and projectile, and $Q_i$ the virtualities of
the hard probes that set the perturbative scale. 

Formula (\ref{sigma}) expresses a factorization property which has been shown
to be equivalent \cite{equivalent} to $k_T$-factorization \cite{kT} in the  
BFKL framework. In this framework, the distributions $\phi^{(i)}(r_i,Q^2_i)$ 
are related to the 
``impact factors'' which describe the coupling of the target and projectile to 
the BFKL kernel. 
In the case of $\gamma^*-$induced reactions, the
dipole distribution functions $\phi^{\g}(r,Q^2)$ are well-known from QED and 
the equivalence 
with photon impact factors checked. In the 
case of forward-jet production with 
transverse momentum $q_T\ge Q\gg 1\,GeV$, the corresponding distribution 
$\phi(r,Q^2)$ can be derived \cite{us,munier} in the collinear approximation 
{\it i.e.} in 
the Double Leading Log approximation (DLL). 

In the present paper, following an approach \cite{motyka} for 
$\gamma^*\!-\!\gamma^*$  
cross-sections, we 
shall describe the predictions of saturation for $g^*-$induced reactions such 
as 
Mueller-Navelet \cite{navelet} forward-jet production. For this sake, we 
will make use of 
the dipole distribution $\phi(r,Q^2)$ derived in \cite{us,munier}. For 
simplicity, we assume 
the validity of the same GBW cross-sections as for $\gamma^*\!-\!\gamma^*$. The 
$k_T$-factorization property is assumed to 
be preserved in the presence of saturation (see a recent 
discussion in  \cite{gelis}). The question of going beyond this simple scheme 
{\it e.g.} using 
a more complete formulation of saturation \cite{kovchegov} is left for further 
work. 

The plan of the paper is as follows. In section II, recalling the results of 
\cite{us,munier}, we show how the emission of a forward gluon jet can be recast 
in terms of a dipole distribution. We also present the GBW formulation 
of the dipole-dipole cross-sections. In section III, we derive our results 
for the Mueller-Navelet jets  
cross-sections with saturation. In 
section IV, we discuss these results in the prospect of experiments at the  
Tevatron and LHC and  
propose  characteristic 
observables for  saturation. The final section 
V is 
devoted to conclusion and  outlook.

\begin{figure}[ht]
\begin{center}
\epsfig{file=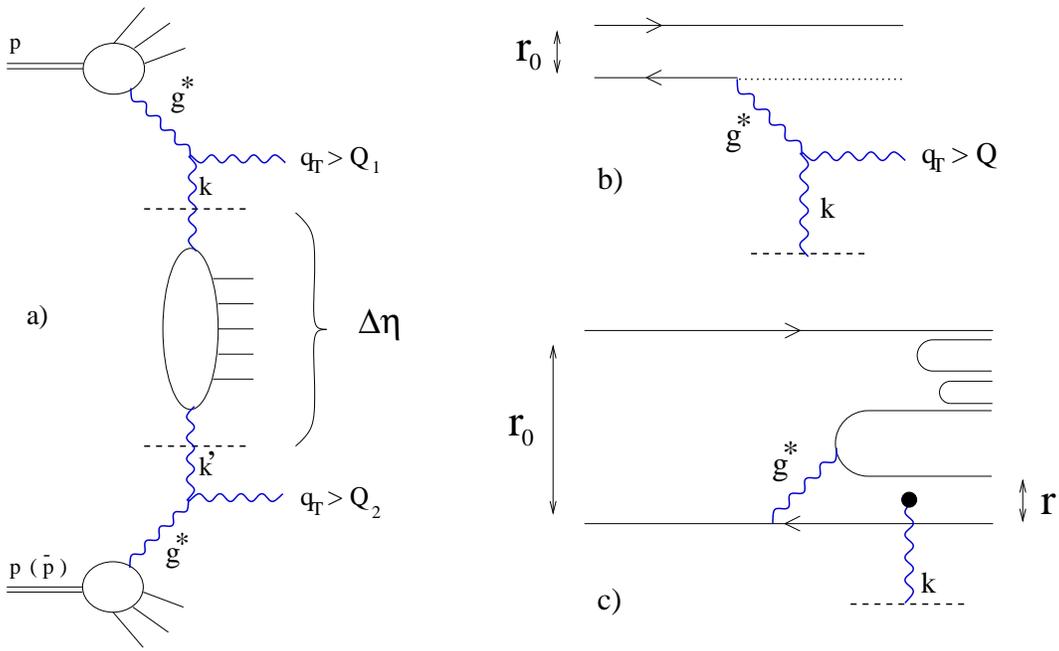,width=14cm}
\caption{{\it Forward jets and impact factors at hadron colliders.} Fig 1a: 
Mueller-Navelet jets at hadron colliders. 
Fig 1b: onium (+jet) impact factor in the partonic representation. Fig 1c: 
onium 
(+jet) impact factor in the dipole representation. $\Delta \eta:$ rapidity 
gap between the two gluon jets.  $q_T>Q_{1,2}:$ tranverse momenta of the gluon 
jets. $k,\ k':$ transverse momenta of the gluons interacting with the BFKL 
kernel. The gluon-dipole coupling ($f^0(k^2,r)$, see text) is sketched by the 
black point in 
Fig.1c.}
\end{center}
\label{F1}
\end{figure}

\section{Dipole formulation}
\label{2}
{\bf A. Forward jets and dipole distributions}

Let us first recall how one can obtain the dipole distribution $\phi(r,Q^2)$  
associated with 
a forward jet with transverse momentum $q_T>Q.$  The derivation is 
made  \cite{munier} using the example of a final-state gluon being emitted from 
an onium ($q\bar q$ state) of size $r_0,$ see Fig1b. QCD factorization will 
allow to extend the result to the case of an incident hadron.

Assuming the condition 
$1\,GeV^{-1}\gg r_0 \gg 1/Q,$ the onium is small 
enough to 
allow a perturbative QCD calculation but large enough with respect to the 
inverse transverse momentum of the forward jet. Using  $k_T$-factorization in 
the BFKL framework
the (unintegrated) gluon density $f(k^2,r_0)$ entering at each vertex of
the BFKL cross-section (see Fig.1b) can be
factorized \cite{munier} in the following way:
\begin{equation}
f(k^2,r_0) \equiv \bar\alpha\log\frac 1x \int \frac{d^2\vec q}{\pi\vec 
q^2}\theta(\vec q^2-Q^2)
f^0(|\vec k+\vec q|^2,r_0)\approx \int_0^{r_0} d^2r\ 
\left\{2\bar\alpha\log\frac 1x
\log\frac{r_0}{r}\right\}\  
\frac Q{2\pi r}\ J_1(Qr)\ f^0(k^2,r) 
\label{phig}
\end{equation}
in the collinear approximation $r_0q_T \gg 1$ for the onium. $k$ is the 
transverse momentum of 
the gluon connected to the BFKL kernel 
(see Fig.1).
$f(k^2,r_0)$ is the lowest order BFKL equation written in an unfolded form (see 
for instance 
\cite{sutton}) and the initial gluon density reads $f^0(k^2,r)\equiv 
2\bar\alpha(1-J_0(kr))/k^2.$ $J_0$ and $J_1$ are the 
Bessel functions.

Equation (\ref{phig}) can be interpreted as the extension to forward jets of 
the equivalence  
\cite{equivalent} between the momentum-space (partonic) and coordinate-space 
(dipole) 
representations. The middle term corresponds to the contribution displayed in 
Fig.1b. The last term is described in Fig.1c and matches with the Mueller 
picture  
\cite{mueller} of cascading dipoles in the $1/N_c$ limit, in which 
the QCD wave 
function of an initial onium is expanded over multi colorless-dipole 
configurations. The factor in brackets $\left\{2\bar\alpha \log\frac 1x
\log\frac{r_0}{r}\right\}$
corresponds to the first order contribution of the  
Dokshitzer-Gribov-Lipatov-Altarelli-Parisi (DGLAP) 
gluon ladder \cite{dglap}, {\it i.e.} the probability of 
finding a dipole of size $r$ inside the onium of size  $r_0,$  at  
the Double Leading Log (DLL) approximation; thanks to QCD factorization 
properties, it is 
included in the gluon structure function of the incident particule.
 $f^0(k^2,r)$ is nothing else than the factorized gluon 
density  \cite{mueller,wallon} inside the dipole of size $r$ which, in the 
dipole formulation (\ref{sigma}) is included in the dipole-dipole 
cross-section.

Having factorized out both the contribution to the structure function and the 
one to the dipole-dipole cross-section, one is left with the function 
$\phi(r,Q^2)$ which describes the resulting size distribution of the 
interacting 
dipole. Hence, one is led to identify\footnote{This formula
can also be obtained  \cite{us} for $\phi(r,z,Q^2)$, taking into account the 
energy fraction $z$ shared between the quark and the antiquark of the 
dipole. However, since 
the dipole-dipole  
cross-sections we will consider are $z-$independent, we only have to consider 
the 
distributions 
$\phi$ integrated over $z.$} 
\begin{equation}
\phi (r,Q^2) \equiv 
\frac Q{2\pi
r}\ 
J_1(Qr)\ .
 \label{bessel}
 \end{equation}
The forward-jet emission is thus put in correspondence with a small colorless 
dipole of size $r = {\cal 
O} 
(1/Q) $.  The distribution of sizes around that value is given by $\phi 
(r,Q^2)$ in (\ref{bessel}).

Some  comments  are in order.  Via its description in terms of   dipoles, 
$k_T$-factorization leads   to a description  of the forward jet (coming from a 
colorful virtual gluon $g^*$) in 
terms of colorless $q\bar 
q$ dipoles. Indeed, within  the 
$1/N_c$  scheme of the dipole formalism, 
the color neutralization of the forward jet is described by a cascade of 
dipoles, as pictured in Fig.1c. 
This means 
that, in this representation, the color quantum number carried by the {\it 
incoming} virtual gluon 
$g^*$ 
is neutralized through the cascade of dipoles.

The obtained
dipole distribution $\phi (r,Q^2)$ is not everywhere positive, the Bessel 
function oscillating in sign for $rQ\gtrsim 4,$
and therefore
cannot be interpreted as a probability 
distribution. We interpret this feature as a breakdown of the collinear 
approximation. Hence,
in our framework, we have to check the positivity of the cross-sections, 
as will be discussed later on. For the Mueller-Navelet BFKL cross-section,  
positivity is satisfied by construction.

\vspace{.3cm}
{\bf B. Dipole-dipole cross-sections with saturation}

Let us recall the formulation of the GBW saturation model for dipole
collisions. Initially, the 
GBW approach \cite{golec} is a model for the dipole-proton cross-section 
which includes the saturation damping of 
large-dipole configurations. For the description of  $\g^*\!-\!\g^*$ 
cross-sections at 
LEP, see Ref. \cite{motyka}, it has been extended\footnote{For our 
purpose, 
we shall only use a  high-energy
approximation of  the expressions quoted by the authors \cite{motyka}.} to 
dipole-dipole cross-sections. The same saturation scale is considered for 
dipole-dipole 
and dipole-proton cross-sections. 
 
The parametrisation of this dipole-dipole cross-section is 
\begin{equation}
\sigma_{dd}(\Delta\eta,r_1,r_2) = \;\sigma_0\left\{
1- \exp\left(-{r_{\rm eff}^2\over  4R_0^2(\Delta \eta)}\right)
\right\},
\label{sigmadd}
\end{equation}
where $R_0(\Delta \eta)=e^{-\f{\lambda}2\left({\Delta 
\eta}-{\Delta \eta}_0\right)}/Q_0$ is the rapidity-dependent
saturation radius and the  dipole-dipole {\it effective} radius 
$r^2_{\rm  eff}(r_1,r_2)$ is defined \cite{motyka} in such a way to satisfy 
{\it color transparency}, namely 
$\sigma_{dd}  
\propto r_{i=1,2}^2$ when  $r_{i} \rightarrow 0.$
  As in \cite{motyka}, three scenarios for $r_{\rm eff}(r_1,r_2)$ will {\it a 
priori} be considered:
\begin{equation}
{\bf 1.}\  \D r^2_{\rm  eff}\; = \;{r_1^2r_2^2\over r_1^2+r_2^2}\hspace{1cm}
{\bf 2.}\  \D r^2_{\rm eff}\; = \;\min(r_1^2,r_2^2)\hspace{1cm}
{\bf 3.}\  \D r^2_{\rm  eff}\; = 
\;\min(r_1^2,r_2^2)\left\{1+\ln\frac{\max(r_1,r_2)}{\min(r_1,r_2)}\right\}\ .
\label{three}
\end{equation}
All three parametrisations exhibit {\it color 
transparency}.
Cases {\bf 1} and {\bf 2} reduce to the original GBW model
when one of the dipoles is much larger than the other
and  the model {\bf 3} corresponds to the dipole-dipole cross-section 
mediated by a two-gluon 
exchange \cite{mueller}. 
For the saturation  radius $R_0(\Delta \eta)$ 
we adopt the same set of parameters\footnote{$\lambda = 
.288,$ $\Delta 
\eta_0= 8.1$ for $Q_0\equiv 1\ GeV.$} as those in
 \cite{golec,motyka}. 
 
\section{Hard cross-sections}
 
Let us derive the general formulae we get for the cross-sections 
(\ref{sigma}). 
Defining $u={r_2}/{r_1}$ and $r^2_{eff}=r^2_1f_i(u),$
the three scenarios considered in (\ref{three}) can be rewritten
\begin{equation}
f_1(u)=\frac{u^2}{1+u^2}\hspace{1cm}f_2(u)=\left\{\begin{array}
{ll} u^2&\mbox{if }u<1\\1&\mbox{if 
}u>1\end{array}\right.\hspace{1cm}f_3(u)=\left\{\begin{array}{ll} 
u^2(1-\log 
u)&\mbox{if }u<1\\1+\log u&\mbox{if }u>1\end{array}\right.\ .
\label{u}
\end{equation}
Then inserting (\ref{bessel}) in
formula (\ref{sigma}) leads to
\begin{eqnarray}\D
\frac{\sigma_i}{\sigma_0}&=&\D 1-Q_1Q_2\int_0^\infty du\int_0^\infty rdr\ 
J_1(rQ_1)J_1(ruQ_2)\ e^{-\frac{r^2}{4R^2_0}f_i(u)}\nonumber\\\D
&=&\D 1-2R^2_0Q_1Q_2\int_0^\infty\frac{du}
{f_i(u)}e^{-(Q^2_1+Q^2_2u^2){R^2_0}{f^{-1}_i(u)}}I_1\left(\frac{2Q_1Q_2uR
^2_0}
{f_i(u)}\right)\ ,
\label{si}
\end{eqnarray}
 after integration over $r.$ 
$I_1$ is the modified Bessel function of the first kind. Formula 
(\ref{si}) gives the 
theoretical cross-sections within the GBW model for hard hadronic probes. 

Let us discuss our results. The dipole distribution 
$\phi(r,Q^2)$ is not everywhere positive but this 
is not {\it a priori} an obstacle as long as the corresponding total 
cross-sections (\ref{sigma}) stay positive. This 
is for instance realized by the  BFKL 
cross-section \cite{us}. It is compulsory to verify whether or not this 
positivity is altered by saturation. 
By numerical inspection of formulae (\ref{si}), we checked that 
the 
positivity constraint is verified. Qualitatively, this is due to the fact 
that the 
negative values of the Bessel functions in (\ref{si}) are present 
for 
large dipole sizes whose contributions are strongly reduced by 
saturation.

Another constraint is to check that the 
cross-sections $\sigma_{dd}\sim\sigma_0\ {r_{\rm eff}^2/  
4R_0^2(\Delta \eta)}$
corresponding to the limit of small dipole sizes in  
(\ref{sigmadd}), lead to cross-sections behaving  like $1/\left\{R_0^2(\Delta 
\eta)\max{(Q_1^2,Q_2^2)}\right\},$ as expected from transparency. 
Computing the 
gluon-gluon cross-section in this  limit gives 
\begin{equation}
\sigma_1 \sim \frac{\sigma_0}{R^2_0}\frac{2Q^2_1Q^2_2}{(Q^2_1+Q^2_2)^3}
\hspace{1cm}\sigma_2 \sim \frac{\sigma_0}{R^2_0}\delta(Q^2_1-Q^2_2)
\hspace{1cm}\sigma_3 \sim 
\frac{\sigma_0}{2R^2_0}\min\left(\frac{1}{Q^2_1},\frac{1}{Q^2_2}\right)
\label{transparency}
\end{equation}
 which shows that the models {\bf 1} and {\bf 3} of (\ref{three}) verify the 
constraint. The model {\bf 2} 
does not, as confirmed by an explicit integration of   
(\ref{si}) 
which gives in this case
\begin{equation}
\frac{\sigma_2}{\sigma_0}=e^{-R_0^2(Q^2_1+Q^2_2)}\ I_0(2R_0^2Q_1Q_2) 
\sim 
\frac {e^{-R_0^2(Q_1-Q_2)^2}}{2R_0\sqrt{\pi Q_1Q_2}}\rightarrow 
\frac{\delta(Q^2_1-Q^2_2)}{R^2_0}\ ,
\label{I0}
\end{equation}
at large $R_0.$ Hence, within our approximations, the model {\bf 2} cannot be 
considered.

It is possible to derive a general formula for 
saturation 
in the dipole framework which could be valid for any hard probe expressed 
in terms 
of the dipole basis, be it a forward jet, an onium, a virtual photon, 
etc... In particular, it will be useful to extend the saturation 
discussion to  forward 
jets at HERA, with a $\g^*\!-\!g^*$ cross-section.
We consider the  Mellin transforms of the dipole distributions $
   \tilde{\phi}(\tau)=\int 
d^2r \ (r^2Q^2)^{\tau}\phi(r,Q^2)\ .$
For instance, one has 
\begin{equation}
\tilde{\phi}(\tau)=4^\tau\frac{\Gamma(1+\tau)}{\Gamma(1-\tau)}
\hspace{1cm}
\tilde{\phi}^\gamma(\tau)\propto\ 
\pi2^{2\tau+1}\frac{\Gamma(1-\tau)\Gamma(3-\tau)\Gamma(\tau)
\Gamma^2(1+\tau)
\Gamma(2+\tau)}{\Gamma(4-2\tau)\Gamma(2+2\tau)}\ .
\label{gg}
\end{equation}
After some straightforward algebra, 
we obtain the following inverse Mellin transform expressions:
\begin{equation}
\frac {\sigma_i}{\sigma_0}=\intc{\tau}\tilde{\phi}^{(1)}(\tau)
(2Q_1R_0)^{-2\tau}\intc{\sigma}\tilde{\phi}^{(2)}(\sigma)
(2Q_2R_0)^{-2\sigma}\ 
g_i(\sigma,\tau)\ ;\hspace{1cm}
0<Re(\sigma),\;Re(\tau),\;Re(\sigma+\tau)<1\ ,
\label{sigene}
\end{equation}
where $\tilde{\phi}^{(1)}$ and $\tilde{\phi}^{(2)}$ are the 
Mellin-transformed dipole 
distributions in the target and projectile and,
 for the different models (\ref{three}), one has
\begin{equation}
\begin{array}{ll}\D 
g_1(\sigma,\tau)=\frac{\Gamma(1-\tau-\sigma)\Gamma(\sigma)
\Gamma(\tau)}
{\Gamma(1+\tau+\sigma)}\hspace{1.5cm}
g_2(\sigma,\tau)=\frac{\Gamma(1-\tau-\sigma)}{\sigma\tau}\\\D 
g_3(\sigma,\tau)=-2^{-\tau-\sigma}\Gamma(-\tau-\sigma)
\left\{e^{2\sigma}
\sigma^{-1-\tau-\sigma}\Gamma(\tau+\sigma+1,2\sigma)+\left[\tau \iff 
\sigma\right] \right\}\ .
\end{array}
\label{g}
\end{equation} This formulation allows us to 
extend easily 
our computations to various cases.
\begin{figure}[ht]
\begin{center}
\epsfig{file=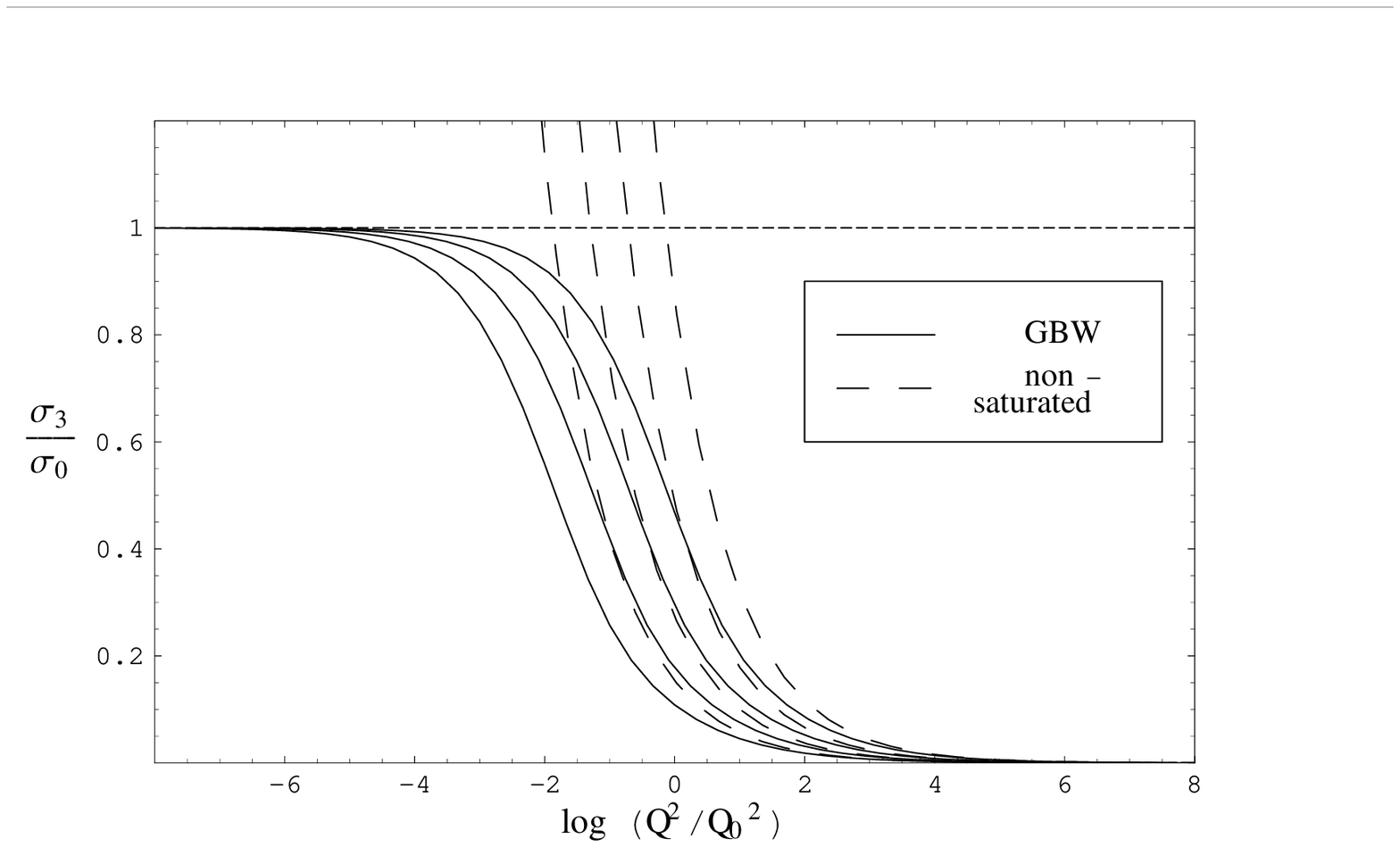,width=17cm,clip=true}
\caption{{\it $g^*\!-\!g^*$   cross-sections (model {\bf 3}).}  $Q_1=Q_2=Q:$ 
symmetric virtuality case. $\Delta \eta =\mbox{(from left to right)} \ 
4,6,8,10:$ rapidity intervals. Full lines: saturation cross-sections 
(\ref{si}). Dashed lines: without saturation (\ref{transparency}).}
\end{center}
\label{F2}
\end{figure}
After easy transformations, one gets for the different GBW models
\begin{equation}
\begin{array}{lllll}\D 
\frac {\sigma_1}{\sigma_0}=\int_0^\infty dx\ J_1(x)\ 
A^{(1)}(xQ_1R_0)A^{(2)}(xQ_2R_0)\ ;\hspace{1cm}A(x)=\intc{\tau}
x^{-2\tau}\tilde{\phi}(\tau)\Gamma(\tau)\\\\\D 
\frac {\sigma_2}{\sigma_0}=\int_0^\infty 2xdx\ e^{-x^2}\ 
B^{(1)}(2xQ_1R_0)B^{(2)}(2xQ_2R_0)\ ; 
\hspace{1cm}B(x)=\intc{\tau}x^{-2\tau}
\frac{\tilde{\phi}(\tau)}{\tau}\\\\\D 
\frac {\sigma_3}{\sigma_0}=\frac {\sigma_2}{\sigma_0}+\int_0^\infty 2xdx\ 
e^{-x^2}\left(C^{(1)}(2xQ_1R_0)D^{(2)}(2Q_2R_0,x)+D^{(1)}(2Q_1R_0,x)C^{(2)}(2
xQ_2R_0)\right)
\end{array}
\label{models}
\end{equation}
where
\begin{equation}
C(x)=\intc{\tau}x^{-2\tau}\tilde{\phi}(\tau)\ 
;\hspace{1cm}D(Q,x)
=\intc{\tau}(Qx)^{-2\tau}\frac{\tilde{\phi}(\tau)}
{\tau(2\tau+x^2)}\ .
\label{CD}
\end{equation}
With these general formulae, we find back our previous results
(\ref{si},\ref{I0}) 
 and obtain\footnote{Note that  the input 
functions $A,B,C,D$ (\ref{models},\ref{CD}) correspond to specific Meijer 
functions.}  those for $\g^*.$

\section{Phenomenological applications}

Let us investigate the phenomenological outcome, for hadron 
colliders, of our extension of the
GBW models to hadronic ({\it i.e.} $g^*$) probes. The theoretical 
cross-sections are obtained from  
formulae
(\ref{u})-(\ref{si}) and (\ref{sigene})-(\ref{CD}), in terms of the physical 
variables $Q_1, Q_2\mbox{ and } 
\Delta 
\eta,$ once the saturation scale parameters $Q_0, \lambda\mbox{ and }\Delta 
\eta_0$
are taken identical to their reference
values (see footnote 3).

In Fig.2, we display the cross-section ratio $\sigma_3/\sigma_0$
as a function of $\log (Q^2/Q^2_0),$ where $ Q=Q_1=Q_2$ for  values of 
the rapidity 
interval $\Delta \eta = 4,6,8,10.$ We also compare with the corresponding 
ratio   for the non-saturated case (\ref{transparency}). As expected, 
the curves show the well-known trend of the GBW model, namely a 
suppression of the non-saturated cross-sections, with a convergence   
towards the full saturation limit $\sigma \rightarrow \sigma_0.$
\begin{figure}[ht]
\begin{center}
\epsfig{file=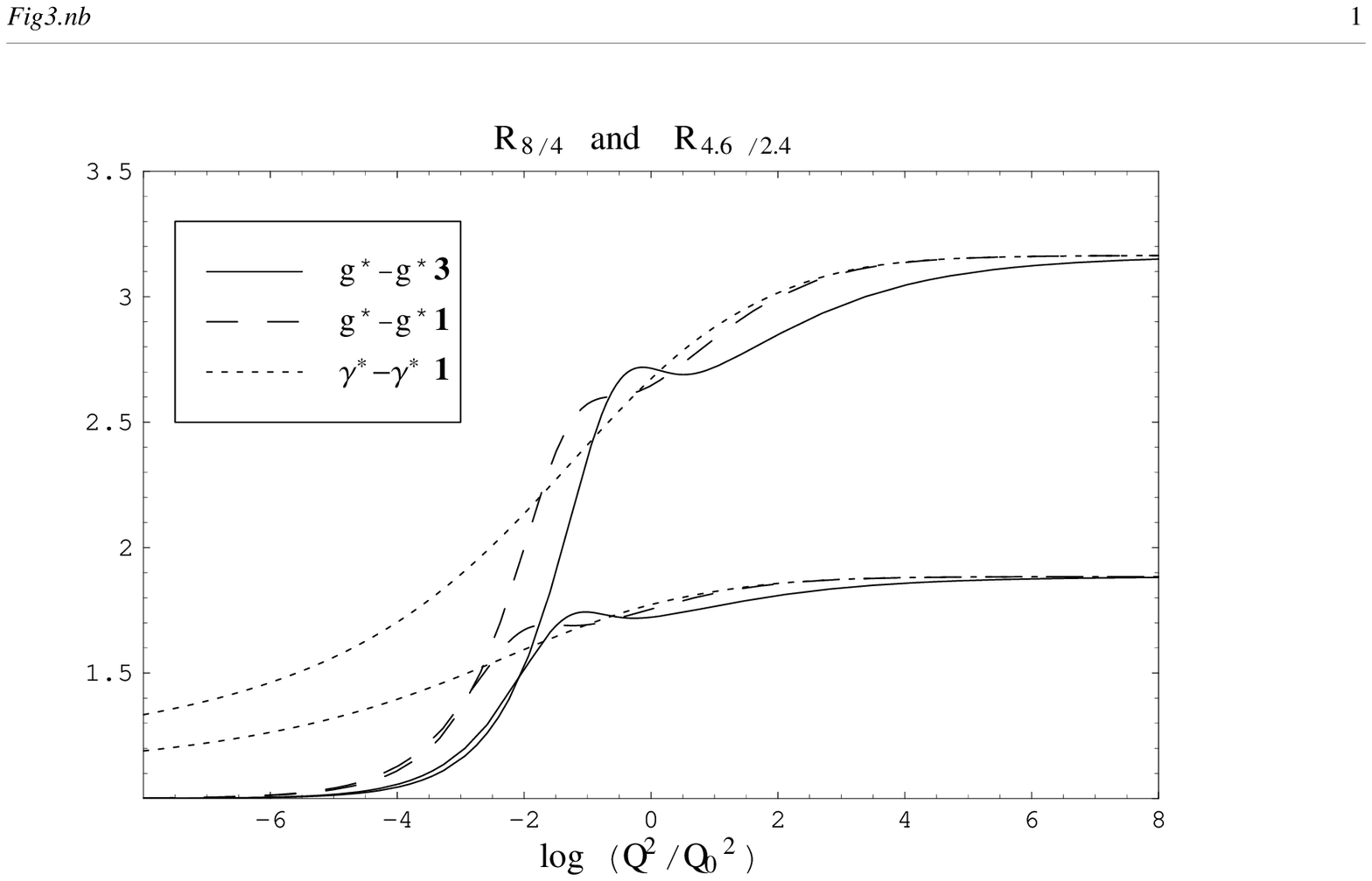,width=17cm,clip=true}
\caption{{\it Cross-section ratios ${\cal R}_{i/j}.$} The resulting ratios 
for models {\bf 1} and  {\bf 3} are plotted for rapidity intervals $i=8,j=4$ 
and $i=4.6,j=2.4$. The comparison is made with $\g^*\!-\!\g^*$ ratios for 
model {\bf 1} and equivalent kinematics. The non-saturated case would be a 
constant  corresponding 
to the high $Q^2$ limit of the plots.}
\end{center}
\label{F3}
\end{figure}
In order to appreciate more quantitatively the influence of saturation, it is 
most convenient to consider the quantities ${\cal R}_{i/j}$ defined as 
\begin{equation}
{\cal R}_{i/j} \equiv  
\frac {\sigma(Q_1,Q_2,{\Delta \eta}_i)}{\sigma(Q_1,Q_2,{\Delta \eta}_j)}\ , 
\label{Rij}
\end{equation}
 {\it i.e.} the cross-section ratios for two 
different values of the rapidity interval. These ratios 
display 
in a clear way the saturation effects.  They  also correspond to possible 
experimental  observables since they can be obtained from measurements at  
fixed values of 
the virtual gluon light-cone momentum and thus are  independent of the 
 gluon structure functions of the incident 
hadrons. Indeed, such observables  have been used for a study of 
Mueller-Navelet jets for testing BFKL predictions at the 
Tevatron \cite{goussiou,jets,jetslhc}.

In Fig.3 we plot the values of ${\cal R}_{4.6/2.4}$ (resp. ${\cal R}_{8/4}$) 
as a 
function of   $Q_1=Q_2\equiv Q.$ These ratios  correspond to values for 
Mueller-Navelet jets studied at the Tevatron \cite{goussiou,jets} (resp. 
realistic 
for the LHC \cite{jetslhc}). The results are displayed both for 
models {\bf 1} and {\bf 3}, see (\ref{three}).
 
As expected from the larger rapidity range, the decrease of ${\cal R}$ 
between the transparency regime and the saturated one is  larger for the LHC 
than for the Tevatron. The striking feature of Fig.3 is that the effect of 
saturation appears as a sharp transition for some critical range in $Q$ 
(higher for the 
LHC).

Let us compare the resulting ratios for hadronic probes ($g^*$-initiated) to 
those for the virtual photon ($\g^*$-initiated) for the same values of the 
rapidity ranges, see Fig.3. Interestingly enough, the photon transition curve 
is 
much smoother, a phenomenon which can be explained by the different structure 
of the
dipole distribution function. Indeed, as discussed 
previously \cite{us}, the dipole distribution  $\phi(r,Q^2)$ has a  tail 
extending
towards large dipole sizes, which are  more damped by the saturation 
corrections. Hence $\phi(r,Q^2)$ is more abruptly cut by 
saturation than the photon dipole distribution $\phi^{\g}(r,Q^2)$. 

In Fig.4, we display the variation of ${\cal R}_{8/4},$ when one looks for 
asymmetric situations, {\it i.e.} $Q_1 > Q_2.$ As seen from the figure, the 
transition may become even sharper in this case, with the formation of a bump 
at a rather 
high value of $Q_1$, which could provide an interesting  
signal for the saturation scale. The origin of this bump lies in the 
different rate of 
increase of the cross-sections towards saturation when the virtualities are 
different. 
This possible signal is present at rather high scale, which could be useful 
for experimental 
considerations, as we shall develop now.

{\bf i) Mueller-Navelet jets}

Two jets  separated by a 
large rapidity interval, or Mueller-Navelet jets \cite{navelet}, are the more 
natural process for  our formulae (\ref{si}) to be applied.
Indeed, a measurement of those dijet cross-sections has been performed at 
Tevatron, with jets of transverse momentum with a lower $E_T$ cut,  related to 
the virtuality $Q.$ To 
actually
mesure the ratio ${\cal R}$ in order to get rid of uncertainties on the
structure functions, the two available incident energies (630 GeV and 
1800 GeV) were used. The result was a strong increase of 
 ${\cal R}$   with the rapidity interval, which was 
pointed out as a possible
hint of BFKL evolution. 
 \begin{figure}[ht]
\begin{center}
\epsfig{file=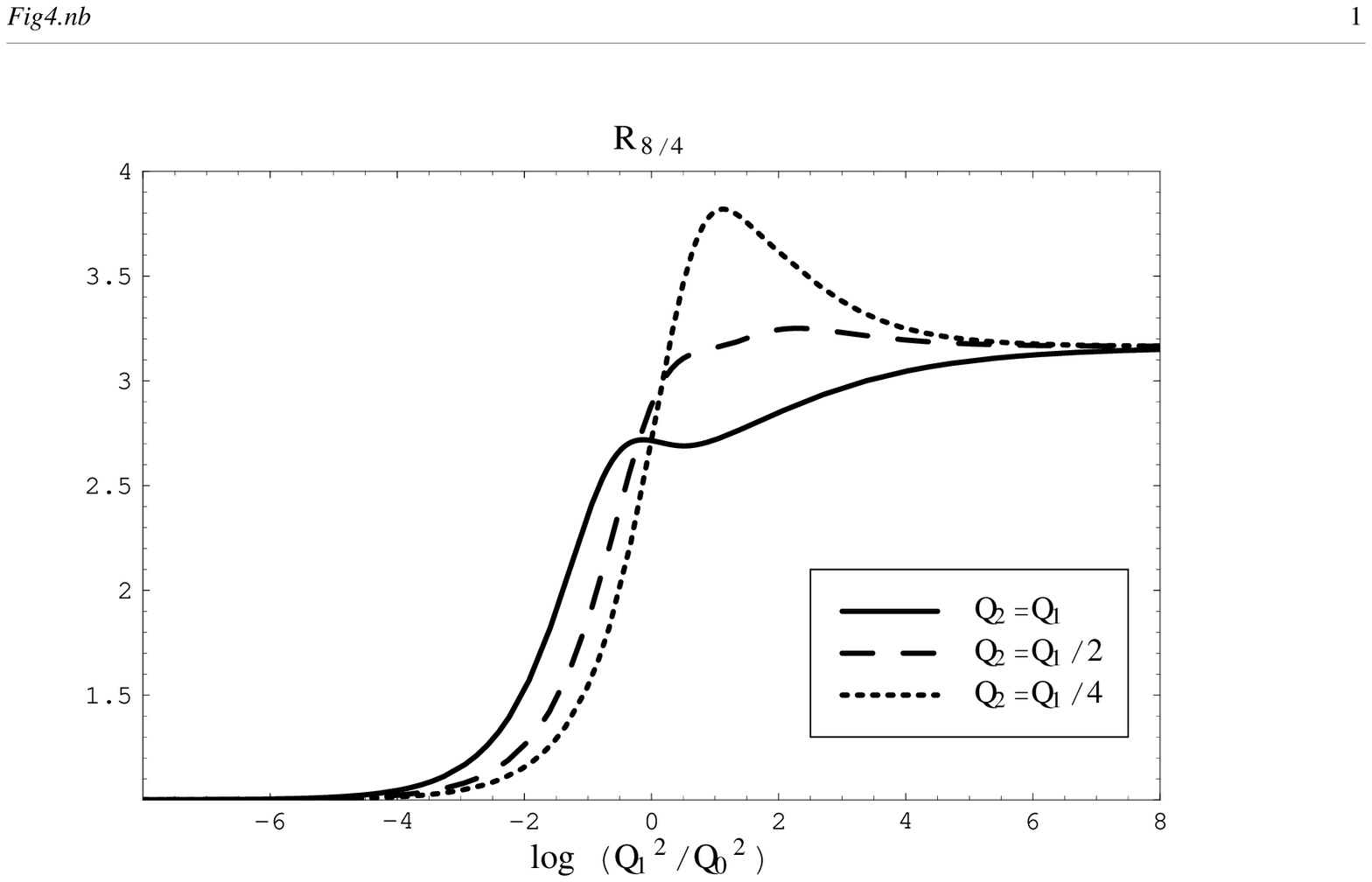,width=17cm,clip=true}
\caption{{\it ${\cal R}_{8/4},$ asymmetric case.} The curves are drawn for 
model {\bf 3}. Note the bump at  $Q_1^2/Q_0^2 \sim 3.$}
\end{center}
\label{F4}
\end{figure}
Saturation studies are favored by  large rapidity intervals (as demonstrated 
in Fig.3). The relevant range of virtuality $Q$ for expecting  a 
clear saturation signal is albeit rather low (see Fig.3, 4). 

If the strong experimental signal reported in  \cite{goussiou} appears to be 
confirmed, the 
saturation prediction displayed in Fig.3 could well be relevant (with a 
redefinition of the 
parameters). However, the  BFKL evolution itself at Tevatron energies 
appears to be quite sizeably modified by finite energy, running 
coupling corrections \cite{stirling} and by experimental cuts \cite{stirling2}. 
Anyway, a simulation of Mueller-Navelet jets that would 
incorporate the 
relation between 
the $E_T$ cut of the jet and the virtuality $Q$ is needed
to discuss the feasability 
of saturation tests in this case. 

{\bf ii) Heavy vector mesons}

As an alternative to hard forward jets, one could consider \cite{royon}  the 
detection 
of two heavy vector mesons with moderate  transverse momentum and separated by 
large rapidity 
intervals. Indeed, 
using $J/\Psi's$ or $\Upsilon's$ may provide a 
hadronic probe of precise mass and transverse momentum. It potentially 
realizes a colorless
$q\bar q$ probe and thus could give an information on the differential 
distribution of dipoles $ \phi(r,Q^2),$ for instance on the dipole-size 
distribution. Moreover, the leptonic decays may facilitate the event 
selection. 

{\bf iii) Charmed and beauty hadrons}

A forward-jet detection corresponds to one of the $q\bar q$ partners of a 
dipole. The detection of a heavy flavored meson    would give a similar  
interesting signal. One thus would look for the detection of heavy flavored  
mesons separated by a large rapidity interval. In particular,  the detection 
of a $D^*$ on one side and a $B$-meson on the other 
side would  realize  interesting asymmetric configurations\footnote {Note that 
asymmetric configurations are preferable to avoid eventual non-BFKL logarithmic 
corrections \cite{stirling2}.} such as seen on 
Fig.4. One could also play with their transverse momentum cuts to vary the 
$g^*$ 
virtualities.

These possibilities of realizing hadronic probes of saturation certainly 
deserve more studies in the near future. Simulations of these processes at 
Tevatron and LHC energies will give a quantitative estimate of the potential 
of hadronic colliders to reveal new features of  saturation.

\section{Conclusion and Outlook}

Let us briefly summarize the main results of our study.

We started from an extension of the  Golec-Biernat and W\"usthoff saturation 
model to hadronic collisions. For 
this sake, we used a QCD dipole formulation \cite{us,munier} of hard hadronic 
probes based on a $k_T$-factorization assumption; such probes as forward 
(Mueller-Navelet) jets, 
heavy mesons, heavy flavoured mesons are initiated by off-mass-shell gluons. 
Our results are:

i) A derivation of saturation predictions for total hard 
cross-sections at hadron colliders, {\it e.g.} the Tevatron and LHC.

ii) Observables which possess a potentially clear signal for 
saturation at high rapidity intervals and gluon virtualities around the 
expected 
saturation scale in  dipole-dipole interactions, see Figs.(2-4).

iii) The suggestion of using, besides the well-known Mueller-Navelet jets,  
heavy vector and/or 
heavy flavored mesons to 
measure hard cross-sections and their transition towards saturation.

There are quite a few open issues for the present formalism:

On a phenomenological ground, it indicates a way for simulations in the 
framework 
of hadron colliders using the QCD dipole formalism which had been so useful 
in 
the HERA context. These phenomenological studies will tell us whether and how 
saturation could be present and checked at the Tevatron and/or 
the 
LHC. Note also that 
forward jets at HERA, which are initiated by $g^*-\g^*$  configurations with 
large rapidity separation, could be interesting to investigate. Beyond the 
scope of the present 
paper, determinant experimental  issues, like fighting against pile-up events, 
background studies, 
possibility of a direct access to the hard cross-section ratios ${\cal 
R}_{i/j},$ etc... deserve to be explored. 

On the theoretical ground, a study going beyond the single $q\bar q$ basis 
for
the hard probe dipole distribution is deserved. In particular, adding the $g 
q\bar q$ or few-dipole configurations is important to discuss the 
$k_T$-factorization assumption. A more complete study of saturation effects in 
the emission of 
an energetic gluon \cite{kovchegov} deserves further investigation. It 
may also allow us to extend 
our formalism to 
diffraction processes, since such configurations appeared important  for the 
GBW model \cite{golec} 
at 
HERA. 
It is also possible to extend the theoretical analysis beyond the GBW 
formulation and to directly introduce 
solutions \cite{iancu} of the non-linear QCD evolution 
equations, which would have a  BFKL (and not merely transparency) limit 
at low density.
\begin{acknowledgments}
We thank Rikard Enberg, Christophe Royon and St\'ephane Munier for  useful 
comments and 
suggestions. 
\end{acknowledgments}


\begin{thebibliography}{10}

\bibitem{bfkl}
L. N. Lipatov, {\it Sov. J. Nucl. Phys.} {\bf 23}, (1976) 338;
E. A. Kuraev, L. N. Lipatov and V. S. Fadin,
{\it  Sov. Phys. JETP} {\bf 45}, (1977) 199;
I. I. Balitsky and L. N. Lipatov,
{\it Sov. J. Nucl. Phys.} {\bf 28}, (1978) 822.

\bibitem{GLR} L. V. Gribov, E. M. Levin and M. G. Ryskin, {\it Phys. Rep.}
{\bf 100}, (1983) 1.

\bibitem{qiu} 
A. H.\ Mueller and J. Qiu, {\it Nucl. Phys.} {\bf B268}, (1986) 427.

\bibitem{venugopalan}
L. McLerran and R. Venugopalan, {\it Phys. Rev.} {\bf D49}, (1994) 2233;
{\it ibid.}, (1994) 3352;
{\it ibid.}, {\bf D50}, (1994) 2225;
A. Kovner, L. McLerran and H. Weigert, {\it Phys. Rev.}
 {\bf D52}, (1995) 6231; {\it ibid.}, (1995) 3809;
R. Venugopalan, {\it Acta Phys. Polon.} {\bf B30}, (1999) 3731;
E. Iancu, A. Leonidov and L. McLerran,
{\it Nucl. Phys.} {\bf A692}, (2001) 583;
{\it Phys. Lett.} {\bf B510}, (2001) 133;
E. Iancu and L. McLerran, {\it Phys. Lett.} {\bf B510}, (2001) 145;
E. Ferreiro, E. Iancu, A. Leonidov and L. McLerran, 
{\it Nucl. Phys.} {\bf A703}, (2002) 489;
H. Weigert, {\it Nucl. Phys.} {\bf A703}, (2002) 823.

\bibitem{Balitsky:1995ub}
I. Balitsky, {\it Nucl. Phys.} {\bf B463}, (1996) 99;
Y. V. Kovchegov,
{\it Phys. Rev.} {\bf D60}, (1999) 034008;
{\it ibid.}, {\bf D61}, (2000) 074018.

\bibitem{levin} 
E. Levin and J. Bartels, {\it Nucl. Phys.} {\bf B387}, (1992) 617;
Y. V. Kovchegov, {\it Phys. Rev.} 
{\bf D61}, (2000) 074018;
E. Levin and K. Tuchin, {\it Nucl. Phys.} {\bf A693}, (2001) 787; 
{\it ibid.},
{\bf A691}, (2001) 779.

\bibitem{Mueller:2002zm}
A. H. Mueller and D. N. Triantafyllopoulos,
{\it Nucl. Phys.} {\bf B640}, (2002) 331.

\bibitem{golec} 
K. Golec-Biernat and M. W\"usthoff, {\it Phys. Rev.} {\bf 
D59} 
(1998) 014017,  {\it Phys. Rev.} {\bf D60} (1999) 114023.
 
\bibitem{dglap}
G. Altarelli and G. Parisi,
{\it Nucl. Phys.} {\bf B126} 18C (1977) 298.
V. N. Gribov and L. N. Lipatov, {\it Sov. Journ. Nucl. Phys.} (1972) 
438 and 675.
Yu. L. Dokshitzer, {\it Sov. Phys. JETP.} {\bf 46} (1977) 641.
For a review: Yu. L. Dokshitzer, V. A. Khoze, A. H. Mueller and
S. I. Troyan 
{\it Basics of perturbative QCD} (Editions Fronti\`eres, J.Tran Thanh Van 
Ed. 1991).

\bibitem{navelet}
A. H. Mueller and H. Navelet, {\it Nucl. Phys.} {\bf B282} (1987) 727.

\bibitem{niko}
N. N. Nikolaev and B. G. Zakharov, {\it Zeit. f\"ur. Phys.} {\bf C49} (1991) 
607;
{\it Phys. Lett.} {\bf B332} (1994) 184.

\bibitem{mueller}
A. H. Mueller, {\it Nucl. Phys.} {\bf B415} (1994) 373;
A. H. Mueller and B. Patel, {\it Nucl. Phys.} {\bf B425} (1994)
471;
A. H. Mueller, {\it Nucl. Phys.} {\bf B437} (1995) 107.
 
\bibitem{equivalent}
S. Munier and R. Peschanski, {\it Nucl. Phys.} {\bf B524 } (1998) 377.
A. Bialas, H. Navelet and R. Peschanski, {\it Nucl. Phys.} {\bf 
B593} (2001) 438.

\bibitem{kT}
S. Catani, M. Ciafaloni and F. Hautmann, {\it Nucl. Phys}. {\bf B366} (1991) 
135.
J. C. Collins and R. K. Ellis, {\it Nucl. Phys.} {\bf B360} (1991) 3;
E. M. Levin, M. G. Ryskin, Yu. M. Shabelsky and A. G. Shuvaev,
{\it Sov. J. Nucl. Phys.} {\bf 53} (1991) 657.

\bibitem{us} R. Peschanski, {\it Mod. Phys. Lett.} {\bf A15} (2000) 1891.

\bibitem{munier}
S. Munier, {\it Phys. Rev.} {\bf D63} (2001) 034015.

\bibitem{motyka}
N. T\^\i mneanu, J. Kwieci\'nski and L. Motyka, {\it Eur. Phys. J.}
 {\bf  C23} 
(2002) 513, {\it Acta 
Phys. Polon.} {\bf 
B33} (2002) 1559 and 3045.

\bibitem{gelis} 
F. G\'elis and R. Venugopalan, {\it Large mass Q-Qbar production from the Color 
Glass Condensate}, 
hep-ph/0310090.

\bibitem{kovchegov}
Yu. V. Kovchegov, {\it Phys. Rev.} {\bf D64} (2001) 114016; Yu. V. Kovchegov, 
K. 
Tuchin, {\it Phys. Rev.} {\bf D65} (2002) 074026.

\bibitem{sutton}
J. Kwiecinski, A. D. Martin and P. J. Sutton, {\it Phys. Rev.} {\bf D52} (1995) 
1445.


\bibitem{wallon}
H. Navelet and S. Wallon, {\it Nucl. Phys.} {\bf B522} (1998) 237.

\bibitem{goussiou} D0 Collaboration: B. Abbott, {\it et al}, {\it 
Phys. Rev. Lett.} 
{\bf 84} (2000) 5722.

\bibitem{jets}
J. G. Contreras, R. Peschanski and C. Royon,
{\it Phys. Rev.} {\bf D62} (2000) 034006.
  
\bibitem{jetslhc}  
R. Peschanski and C. Royon, {\it  Pomeron intercepts at colliders}, 
Workshop on physics at LHC,  hep-ph/0002057.


\bibitem{stirling}
L. H. Orr and W. J. Stirling, {\it Phys. Lett.} {\bf B429} 
(1998) 135.

\bibitem{stirling2} J. R. Andersen, V. Del Duca, S. Frixione, C. Schmidt and
W. J. Stirling, {\it JHEP} {\bf 0102} (2001) 007.

\bibitem{royon}
C. Royon, private communication.

\bibitem{iancu} 
E. Iancu, K. Itakura and S. Munier,
{\it Saturation and BFKL dynamics in the HERA data at small x},
hep-ph/0310338.

\end{thebibliography}
\end{document}